\documentclass[aps,prb,twocolumn,superscriptaddress,floatfix,showpacs]{revtex4}
\usepackage{graphicx}
\usepackage{soul}
\usepackage{amssymb}
\usepackage{amsmath}

\newcommand{\ssm}{\scriptscriptstyle\rm}

\renewcommand{\phi}{\varphi}

\newenvironment{spinor}
{%
\left\{\!\!
\begin{array}{c}
}
{
\!\!
\end{array}
\!\!\right\}
}

\newcommand{\bfpsi}{\boldsymbol \psi}

\begin{document}

\title{Mesoscopic Aspects of Strongly Interacting Cold Atoms}
\author{S.D.\ Huber}
\affiliation{Theoretische Physik, ETH Zurich, CH-8093
Z\"urich, Switzerland}
\affiliation{Department of Condensed Matter Physics, The Weizmann
Institute of Science, Rehovot 76100, Israel}
\author{G.\ Blatter}
\affiliation{Theoretische Physik, ETH Zurich, CH-8093
Z\"urich, Switzerland}

\date{\today}

\begin{abstract}
Harmonically trapped lattice bosons with strong repulsive interactions
exhibit a superfluid-Mott-insulator heterostructure in the form of a
``wedding cake''.  We discuss the mesoscopic aspects of such a system within
a one-dimensional scattering matrix approach and calculate the scattering
properties of quasi-particles at a superfluid-Mott-insulator interface as
an elementary building block to describe transport phenomena across such a
boundary. We apply the formalism to determine the heat conductivity through
a Mott layer, a quantity relevant to describe thermalization processes in
the optical lattice setup. We identify a critical hopping below which the
heat conductivity is strongly suppressed.
\end{abstract}


\maketitle

\section{Introduction}

Cold bosonic gases subject to an optical lattice allow for an accurate
emulation of the Bose-Hubbard model of interacting lattice bosons,
\cite{Jaksch98} with a short range (on-site) interaction between bosons,
a hopping restricted to nearest-neighbors, and no inter-band transitions,
at least for sufficiently deep lattice potentials. Furthermore, the system
is almost perfectly decoupled from environmental degrees of freedom during
typical experimental times. These aspects make cold atoms a perfect
test bed for probing lattice Hamiltonians relevant to condensed matter
physics.\cite{Lewenstein07}

\begin{figure}[t]
\includegraphics{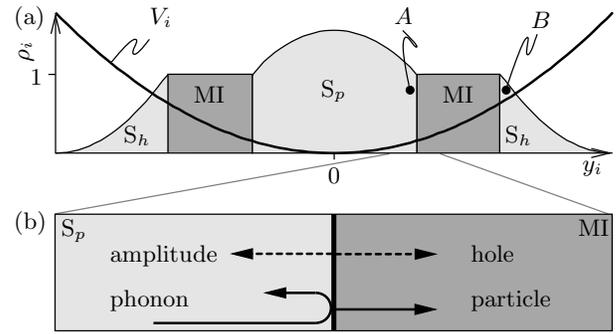}
\caption{
(a) One-dimensional model of a superfluid-Mott-insulator (S-MI)
heterostructure in a trap defined by the potential $V_{i}$ at position $i$.
The dark areas denote Mott-insulating regions with a fixed density $\rho_i$,
while the bright areas are superfluids of condensed particles (center,
S$_{p}$) and condensed holes (wings, S$_{h}$). (b) Sketch of a scattering
event at the superfluid--Mott-insulator interface, where a phonon incident
from the superfluid generates back-reflected quasi-particles of phonon- and
amplitude-type as well as particle- and hole-type excitations transmitted
into the Mott insulator.
}
\label{fig:overview}
\end{figure}

There is one feature, however, where cold atom systems differ from the
generic solid-state setup, as they typically experience an inhomogeneous
potential due to the trap, and thus inferring bulk properties is often
hampered by finite size effects.\cite{Wessel04,Gygi06} On the other hand,
as a result of this confinement, new interesting structures and effects
may occur, with the ``wedding cake'' involving layers of Mott-insulating
and superfluid phases of strongly correlated bosons providing a prominent
example\cite{Kashurnikov02, Folling06} [cf. Fig.~\ref{fig:overview}(a)]. The
evolution of the ground state (gs) across this inhomogeneous system
can be easily understood within the framework of a local density
approximation combined with a mapping between the position in the trap
and the corresponding point in the bulk phase diagram.\cite{fisher89}
The most pronounced new feature in this layered structure is the
superfluid-Mott-insulator (S-MI) interface.  Such boundaries between phases
with different symmetries are known to exhibit interesting effects; a well
known example is the phenomenon of Andreev reflection \cite{Andreev64}
between a normal-metal and a superconductor, where electrons incident on
the superconductor from the normal metal are reflected back in the form
of a hole retracing the electron's path.  Given the strongly correlated
nature of the two phases framing the S-MI interface, similar interesting
phenomena may be expected in the present case. In this work, we provide
a description of the scattering properties of elementary excitations at
the S-MI interface. Such knowledge then allows us to determine the energy
(or heat) flow across interfaces and we determine the heat conductivity
across the Mott insulator region connecting two superfluids.

Mesoscopic aspects in a wedding cake structure have been studied by
Vishveshwara and Lannert, \cite{Vishveshwara08} who calculated the
Josephson coupling across a Mott-insulator domain (a S-MI-S junction)
supported by the overlap between exponentially suppressed ground state
wave functions (superfluid order parameters) across the Mott phase. Here,
we are interested in the scattering dynamics of the excitations near a S-MI
boundary and the transport associated with them across one or multiple
interfaces; these excitations are sound (Goldstone) and massive (Higgs)
modes within the superfluid phase, and particle- and hole-type excitations
in the Mott insulator.\cite{Huber07}

After developing the general framework describing transport in an
inhomogeneous system with phase boundaries, we calculate analytically
the reflection and transmission coefficients for a phonon mode of the
superfluid incident on a Mott insulator [cf.\ Fig.~\ref{fig:overview}(b)].
While these coefficients can be used to describe the scattering of a
wave-packet of phonons (a density disturbance), here they mainly serve as
an illustration of our approach. We then use our results (including those
for scattering of massive modes) to calculate the heat transport through
a Mott barrier.  Knowledge of the heat conductivity $\kappa$ allows to
estimate the thermal contact between superfluid shells in the wedding
cake structure of a trapped Bose system. Note that temperature (actually
entropy) imbalances quite naturally occur in optical lattice systems as
the lattice potential is turned on and entropy is expelled from the newly
formed Mott-insulating regions.\cite{Pollet08} Our heat conductivity $\kappa$
then relates to the thermalization process across different superfluid rings.

Before developing our formalism in detail, we give an overview of the
ideas and concepts utilized in this paper. The excitations close to one of
the Mott-insulating lobes involve coherent superpositions of various site
occupation numbers, and their wave functions are given by a four-spinor
structure. This is reminiscent of the two-spinor structure of excitations
in the Bogoliubov--de Gennes equations describing an inhomogeneous
superconductor (note that for most unconventional superconductors, the
reduction to a two-spinor is possible \cite{Honerkamp98}). The program to be
carried out in order to find the transmission, reflection, and transformation
of quasi-particles then is identical to the one introduced by Blonder,
{\em et al.} \cite{Blonder82} for the superconductor-normal-metal boundary
(in order to simplify the analysis, we consider here a one-dimensional
situation and leave geometric effects due to finite impact angles for
a later study): First, we determine the excitation energies $\hbar
\omega_{X}(k)$ and the four-spinor structure ${\bf X}(k)$ for the
quasi-particle excitations in the superfluid (${\bf X}={\bf s}, {\bf m}$:
sound and massive modes) and in the Mott-insulator (${\bf X}={\bf p}, {\bf
h}$: particle and hole modes); below, this will be done for a translation
invariant situation in a second-quantized formalism. Second, we switch to
a first-quantized formulation and account for the inhomogeneous setup;
for a slowly varying potential [$V_i \to V(y)$] this can be done within
a quasi-classical\cite{Migdal77} or Wentzel-Kramers-Brillouin (WKB)
\cite{Wentzel26,Kramers26,Brillouin26} approximation, with the wave
functions assuming the form
\begin{equation} 
\label{eqn:qcwf}
  \bfpsi_{X}^{\epsilon}(y)=\frac{1}{\sqrt{|k_{X}^{\epsilon}(y)|}}
  \exp\Bigl[i\int^y dx\,
  k_{X}^{\epsilon}(x)\Bigr] {\bf X}[k_{X}^{\epsilon}(y)].
\end{equation}
For a quasi-particle with energy $\epsilon$, the wave vector
$k_{X}^{\epsilon} (y)$ is obtained via proper inversion of the dispersion
$\hbar \omega_{X}(k) [V(y)] = \epsilon$, where the potential $V(y)$ enters
the expression via a shift of the chemical potential, $\delta \mu \to
\delta \mu - V(y)$.  Third, we account for different phases in the setup
by calculating the transfer matrix across the interface \cite{Azbel83} via
matching of the wave functions and their derivatives at the boundary. For
the case of a phonon incident from a (particle-type) superfluid S$_p$ on
a Mott-insulator MI [cf.\ Fig.~\ref{fig:overview}(b)], the wave functions
locally assume the form
\begin{align}
\nonumber
   \bfpsi_{\ssm S}^{\epsilon}(y)
      &=\bfpsi_{s}^{k_{s}^{\epsilon}}(y)
               +r_{\ssm ss} \bfpsi_{s}^{-k_{s}^{\epsilon}}(y)
               +r_{ms} \bfpsi_{m}^{-k_{m}^{\epsilon}}(y), \\
   \bfpsi_{\ssm MI}^{\epsilon}(y)
      &= \tau_{ps} \bfpsi_{p}^{k_{p}^{\epsilon}}(y) 
       + \tau_{hs} \bfpsi_{h}^{k_{h}^{\epsilon}}(y),
   \label{eqn:swf}
\end{align}
with $\bfpsi_{X}^{\pm k}(y)$ denoting plane-wave functions of type
(\ref{eqn:qcwf}) with constant wave vector $\pm k$. The scattering amplitudes
$\tau_{ps}$ (transmitted particle), $\tau_{hs}$ (transmitted hole), $r_{ss}$
(reflected sound), and $r_{ms}$ (reflected massive mode) are obtained
from the continuity conditions across the interface, $\bfpsi_{\ssm
S}^{\epsilon}(0)=\bfpsi_{\ssm MI}^{\epsilon}(0)$ and $\partial_{y}
\bfpsi_{\ssm S}^{\epsilon}(0)=\partial_{y} \bfpsi_{\ssm MI}^{\epsilon}(0)$.

In order to fully characterize the S$_p$-MI boundary, one has to
determine the $4 \times 4$ transfer matrix relating phonon and massive
modes propagating to the left and right through the superfluid S$_p$ with
the particle and hole modes propagating to the left and right through the
Mott-insulator, requiring the solution of four scattering problems of the
above type [Eq. (\ref{eqn:swf})] (the corresponding task has to be solved
in order to describe the interface between a hole-type superfluid S$_{h}$
and a Mott insulator).  In the following, we concentrate exclusively on the
scattering properties at the interface where the superfluid order parameter
$\psi$ vanishes; this guarantees the solvability of the matching conditions
at the boundary. As our final result, we present: (i) simple analytical
expressions for the scattering coefficients at small hopping $t \to 0$
and near the critical value $t_c$ at the tip of the Mott-insulator lobe, as
well as numerical results for two cases in between; and (ii) the behavior
of the heat conductivity $\kappa$ as a function of the model parameters
and the temperature $T$.

The body of the paper is organized as follows: In Sec.~\ref{sec:formalism},
we derive the elementary excitations in the vicinity of the Mott-insulating
regions of the Bose-Hubbard model. We briefly review the mapping to a
\mbox{spin-1} problem \cite{Huber07} before discussing the spinor structure
and symmetry properties of the wave functions. Section~\ref{sec:scatcoff}
is devoted to the calculation of the scattering coefficients for a
superfluid-Mott-insulator interface, and the heat conductivity through a
Mott barrier is determined in Sec.\ \ref{sec:heatcond}. We summarize our
results and conclude in Sec.~\ref{sec:conclusions}.

\section{Excitations}
\label{sec:formalism}

In this section, we first derive the excitations of the strongly
correlated superfluid and the Mott-insulating phase and exploit the
time-reversal symmetry of the problem to obtain a well-suited set of spinor
wave functions.  For the reader who is less interested in the technical
details, the energies $\hbar \omega_{X}(k)$ in Eqs.\ (\ref{eqn:deltapart})
and (\ref{eqn:deltamass}), the spinor wave functions ${\bf X}(k)$ in Eqs.\
(\ref{eqn:mottev}) and (\ref{eqn:sfev}), and their time-reversal invariant
combinations [Eq.\ (\ref{eqn:t-states})], represent the main result of this
section and Table~\ref{tab:spinor} provides a physical interpretation of
the four-spinor components.

We find the dispersion and eigenfunctions of quasiparticle excitations
of the Bose Hubbard-model following the procedure described in detail in
Ref.~\onlinecite{Huber07}. Starting with the Bose-Hubbard Hamiltonian (the
bosonic operators $a_{i}$ create particles in Wannier states at site $i$,
$\delta n_{i}=a_{i}^{\dag}a_{i}-n_{0}$ measures the density from a mean
density $n_{0}$, $t$ is the hopping energy, $U$ accounts for the on-site
interaction, $\delta\mu$ is the chemical potential controlling the particle
number, and $V_{i}$ accounts for the harmonic confinement),
\begin{equation}
\label{eqn:bh}
H=
-t\sum_{\langle i,j \rangle}a_{i}^{\dag}a_{j} 
+\frac{U}{2}\sum_{i}(\delta n_{i})^{2}-\sum_{i} (\delta\mu-V_{i}) \delta n_{i},
\end{equation}
we first consider a homogeneous situation with $V_{i}\equiv 0$
and truncate the bosonic Hilbert space to a site basis with three
local states $|0\rangle_{i}$, $|\!\pm\! 1 \rangle_{i}$.  The state
$|0\rangle_{i}$ refers to $n_{0}$ bosons on site $i$, while the states
$|\!\pm\! 1\rangle_{i}$ include one more (less) particle.  Within this
restricted space, we assume a trial wave function for the ground state
$|\Psi\rangle=\prod_{i}|\psi_{i}\rangle$, where
\begin{equation}
\nonumber
  |\psi_{i}\rangle = \cos(\vartheta)|0\rangle_{i} + \sin(\vartheta)
   \bigl[ \sin(\chi) |\!+\!1\rangle_{i} +\cos(\chi) |\!-\!1\rangle_{i} \bigr].
\end{equation}
Minimizing the variational energy $\varepsilon_{\ssm var}=\langle \Psi|
H_{\ssm BH}| \Psi\rangle$ with respect to $\sigma=\pi/4-\chi$ and
subsequently expanding $\varepsilon_{\ssm var}$ in the order parameter
$\psi=\langle \Psi|a_{i}|\Psi\rangle=\sqrt{n_{0}/2}\sin(2\vartheta)$,
we obtain $\varepsilon_{\ssm var}\approx a\psi^{2}+b\psi^{4}/2$. To
simplify expressions, we assume large filling with $\sqrt{n_{0}+1}\approx
\sqrt{n_{0}}$.  With the critical hopping $t_{c}=U/8n_{0}$, we find for
the order parameter close to the upper (S$_p$-MI) and lower (S$_{h}$-MI)
phase boundaries $\delta\mu_{c}^{\pm}=\pm U \sqrt{1-t/t_{c}}/2$ the
result $|\psi|^{2}= \sqrt{1-t/t_{c}}\; \Delta^{\pm}_{\delta\mu}/t$, where
$\Delta^{\pm}_{\delta\mu}= \delta\mu_{c}^{\pm} \pm \delta\mu$.

The calculation of the excitations above the mean-field ground state
$|\Psi\rangle$ involves a spin-wave analysis in a slave boson description;
the ground state in this effective pseudo-spin 1 formalism defines the
direction of the magnetic order, while the remaining two degrees of freedom
provide the excitations. For $\psi \to 0$, the result can be expressed as a
$4\!\times\!4$ matrix Hamiltonian $H = \sum_{mn} T_m^\dagger {\cal H}_{mn}
T_n$ with four-spinor operators ${\bf T} =(t_{1,k}^{\dag}, t_{-1,k}^{\dag},
t_{1,-k}, t_{-1,-k})$ describing particle and hole creation and destruction
above a mean-filling $n_0$ (cf.\ Table~\ref{tab:spinor}) (the operators
$t_{\pm 1,i}^{\dagger}$ act on the vacuum states $|\mathrm{vac}\rangle_i$
and relate to the bosonic operators $a_i^\dagger$ via $t_{\pm 1,i}^{\dag} =
(a_{i}^{\dagger})^{n_{0}\pm 1}/\sqrt{(n_{0}\pm 1)!}$).

The diagonalization via a Bogoliubov transformation provides the dispersions
(we measure wave vectors $k$ in units of $1/a$, with $a$ the lattice
constant)
\begin{equation} \label{eqn:deltapart}
   \hbar \omega_{p(h)}(k) = \frac{U}{2}
   \sqrt{1-(t/t_{c})\cos(k)}\mp\delta\mu
\end{equation}
in the Mott-insulator and
\begin{equation} \label{eqn:deltamass}
   \hbar\omega_{s(m)}(k)= 4tn_{0}\sqrt{\alpha_{s(m)}^{2}-\beta_{s(m)}^{2}}
    \stackrel{k\rightarrow 0}{\approx}
    \begin{cases} ck,   \\
               \Delta_{m}
    \end{cases}
\end{equation}
in the superfluid phase for $\psi\rightarrow 0$ at $t<t_{c}$, where
\begin{eqnarray*}
   \alpha_{s}&=& [2/(2-t/t_{c})-\cos(k)]/2, 
   \qquad\alpha_{m}= 2t_{c}/t+\alpha_{s}, \\
   \beta_{s}&=& -\beta_{m} =t \cos(k) /(2t_c-t).
\end{eqnarray*}
The sound velocity $c$ of the phonon is $c=\sqrt{4Utn_{0}}$ $/(2t_{c}/t-1)$
and the gap $\Delta_m$ of the massive (or amplitude) mode is given by
$\Delta_{m}=U\sqrt{1-t/t_{c}}$. At the S$_p$-MI interface, the particle
excitation in the Mott-insulator transforms into the (particle-type) sound
mode in S$_p$, while the hole excitation in the Mott-insulator transforms
to the (hole-type) massive mode in S$_p$.  Correspondingly, the bottom of
the hole branch in the Mott-insulator matches up with the gap $\Delta_{m}$
of the massive (hole-type) mode in S$_p$, while the bottom of the particle
branch in the Mott-insulator goes to zero. The use of Eq.~(\ref{eqn:qcwf})
within an inhomogeneous superfluid phase requires generalization of the
result in Eq.~(\ref{eqn:deltamass}) to finite values of $\psi$, which
provides the dependence on $\delta \mu$ away from $\delta \mu_c^\pm$
and hence on the local potential $V(y)$. Within the Mott region, the
particle- and hole spectra undergo a simple shift $\pm \delta\mu$ [ cf.\
Eq.\ (\ref{eqn:deltapart})], with a corresponding simple dependence on
the smooth potential $V(y)$.

The spinor eigenstates in the Mott insulator are
\begin{equation}
\label{eqn:mottev}
{\bf p}_{k}=
\begin{spinor}
A(k) \\0 \\0 \\B(k)
\end{spinor}
\qquad \mbox{and} \qquad
{\bf h}_{k}=
\begin{spinor}
0\\ -A(k) \\-B(k) \\0
\end{spinor},
\end{equation}
with the coefficients
\begin{align*}
A(k) &=\cosh({\rm atanh}\{(t/t_{c})\cos(k)/[2\!-\!(t/t_{c})\cos(k)]\}/2),\\
B(k) &=\sinh({\rm atanh}\{(t/t_{c})\cos(k)/[2\!-\!(t/t_{c})\cos(k)]\}/2),
\end{align*}
providing the ``dressing'' of a particle (with amplitude $A$ in ${\bf
p}_{k}$) by missing hole-type fluctuations (with amplitude $B$) in
the ground state (and vice versa for ${\bf h}_{k}$) and fulfill the
normalization condition.
\begin{table}[t] \begin{center} \begin{tabular}{c|l|l} four-spinor & Microscopic
interpretation& Dirac\\ \hline\hline &&\\[-3pt] $ \begin{spinor}
t_{1,k}^{\dag} \\ t_{-1,k}^{\dag} \\ t_{1,-k} \\ t_{-1,-k}\phantom{,}
\end{spinor} $ & $ \begin{array}{l} \mbox{create particle}\\ \mbox{create hole} \\
\mbox{destroy particle fluctuation in gs} \\ \mbox{destroy hole fluctuation 
in gs} \end{array} $ & $ \begin{array}{l} \mbox{$e^{-}$, $\uparrow$}\\
\mbox{$e^{-}$, $\downarrow$} \\ \mbox{$e^{+}$, $\uparrow$}\\ \mbox{$e^{+}$,
$\downarrow$} \end{array} $ \end{tabular} \end{center} \caption[4-Spinor]{
Physical interpretation of the four-spinor and its similarity to the
Dirac spinor.  In both cases, we deal with two species of ``particles'':
the particle and hole excitations discussed here correspond to spin-up/down
electrons in the Dirac equation. The role of anti-particles in the Dirac
spinor is played by ground state fluctuations in the present context:
the anti-particle of a particle is a fluctuation with a ``missing''
particle in the gs and similar for the hole. Hence the top and bottom
entries are both particle-type (creation of particle and destruction of
a hole), while the two middle entries are both hole-type (creation of a
hole and destruction of a particle).
}
\label{tab:spinor}
\end{table}

The spinor eigenstates in the particle-condensed superfluid phase S$_{\ssm
p}$ are
\begin{equation}
\label{eqn:sfev}
   {\bf s}_{k}^{\ssm p} 
   = \begin{spinor}
     -\Omega_{+}A_{s}(k)\\ \Omega_{-}A_{s}(k)\\ -\Omega_{+}B_{s}(k) \\
        \Omega_{-}B_{s}(k)
     \end{spinor}
   \quad \mbox{and} \quad
   {\bf m}_{k}^{\ssm p}
   = \begin{spinor}
     \Omega_{-}A_{m}(k)\\ \Omega_{+}A_{m}(k)\\ \Omega_{-}B_{m}(k) \\
     \Omega_{+}B_{m}(k)
     \end{spinor},
\end{equation}
with the coefficients
\begin{align}
A_{s(m)} & =
\cosh\{{\rm atanh}[\beta_{s(m)}(k)/\alpha_{s(m)}(k)]/2\}, \\
B_{s(m)} & =
\sinh\{{\rm atanh}[\beta_{s(m)}(k)/\alpha_{s(m)}(k)]/2\}.
\end{align}
Furthermore, $\Omega_{\pm}= [\cos(f(t))\pm \sin(f(t))]/\sqrt{2}$ with
$f(t)=\arctan [2(t_{c}/t)\sqrt{1-t/t_{c}}]$. The eigenstates for S$_{h}$
are obtained by replacing
\begin{equation}
\label{eqn:replacement}
\Omega_{+}\leftrightarrow \Omega_{-}.
\end{equation}

The nature of these excitations is easily understood in the limits $t \to 0$
and $t \to t_c$. For $t \to 0$, the coefficients $A, A_{s(m)} \to 1$ and $B,
B_{s(m)} \to 0$, telling us that the ground state turns into a classical
one devoid of fluctuations  [cf.\ Table\ \ref{tab:spinor}]. The excitations
[Eq. (\ref{eqn:mottev})] in the Mott phase combine particles with absent hole
fluctuations and holes with absent particle fluctuations and hence are of
particle and hole types, respectively. With $\Omega_+ \to 1$ and $\Omega_-
\to 0$ the weights are set differently in the superfluid phase; here, the
excitations involve particles dressed with holes and holes dressed with
particles, reflecting the collective nature of these excitations. Crossing
the boundary $\delta \mu^+_c$ into the particle-type superfluid S$_p$, the
particle mode of the Mott insulator condenses, imprinting the predominant
particle nature onto the sound mode [the gapped hole mode goes over to
the massive (amplitude) mode in the superfluid].

For $t \to t_c$,  the coefficients $A, A_{s(m)} \to \infty$ and $B, B_{s(m)}
\to \infty$; the excitations exploit the strong fluctuations in the ground
state but keep their particle and hole character in the Mott insulator.
With $\Omega_\pm \to 1/\sqrt{2}$, both sound and massive modes have a
mixed particle-hole character and draw large weights from the ground-state
fluctuations.

In the next section, we will have to match the spinor wave functions and
their derivatives, providing us with eight conditions for only four unknown
scattering coefficients. In order to eliminate the additional spurious
conditions, it is convenient to exploit the symmetry under time reversal
and we introduce the time-reversal operator
\begin{equation}
   {\mathcal T} = {\mathcal C}
   \begin{pmatrix} 0 & \sigma_{0} \\ \sigma_{0} & 0 \end{pmatrix},
\end{equation}
where $\mathcal C$ denotes the action of complex conjugation and $\sigma_{0}$
is the identity matrix in two dimensions. The form of ${\mathcal T}$
can be inferred from the behavior of $a_{i}$ under time reversal and
the relations between the operators $a_i$ and $t_{i,0}, t_{i,\pm 1}$. In
addition to the above spinors ${\bf X} = {\bf p},{\bf h},{\bf s},{\bf m}$,
we define the four linearly independent eigenvectors ${\mathcal T} {\bf
X}$. For the situation with unbroken ${\mathcal T}$ invariance discussed
here,\cite{note:josephson} it is convenient to define the new states,
\begin{equation}
   {\bf X}_{k}^{\pm} = ({\bf X}_{k}\mp i {\mathcal T}{\bf X}_{k})/\sqrt{2}.
\label{eqn:t-states}
\end{equation}
These form a basis of ${\mathcal T}$ representations and are multiplied with
$\pm i$ under the operation ${\mathcal T}$.  Without loss of generality,
we choose to work with the ``+''-eigenstates and drop the superscript in
the following.

With these spinors at hand, we are now in the position to describe
inhomogeneities.  For piecewise constant ``potentials'', we can use
plane-wave-type left and right moving spinor wave functions
\begin{equation}
\label{eqn:wavefunctions}
   \bfpsi_{X}^{\pm k_{X}^{\epsilon}}(y)=e^{i(\pi\mp\pi)/4}
   e^{\pm ik_{X}^{\epsilon} y}{\bf X}_{k_{X}^{\epsilon}},
\end{equation}
and combine them into the scattering states defined in Eq.\
(\ref{eqn:swf}).  For slowly varying parameters (``potentials''), the
states [Eq.\ (\ref{eqn:t-states})] determine the spinor structure in Eq.\
(\ref{eqn:qcwf}).

\section{Scattering coefficients}
\label{sec:scatcoff}

\begin{figure*}[t]
\begin{center}
\includegraphics[width=17.5cm]{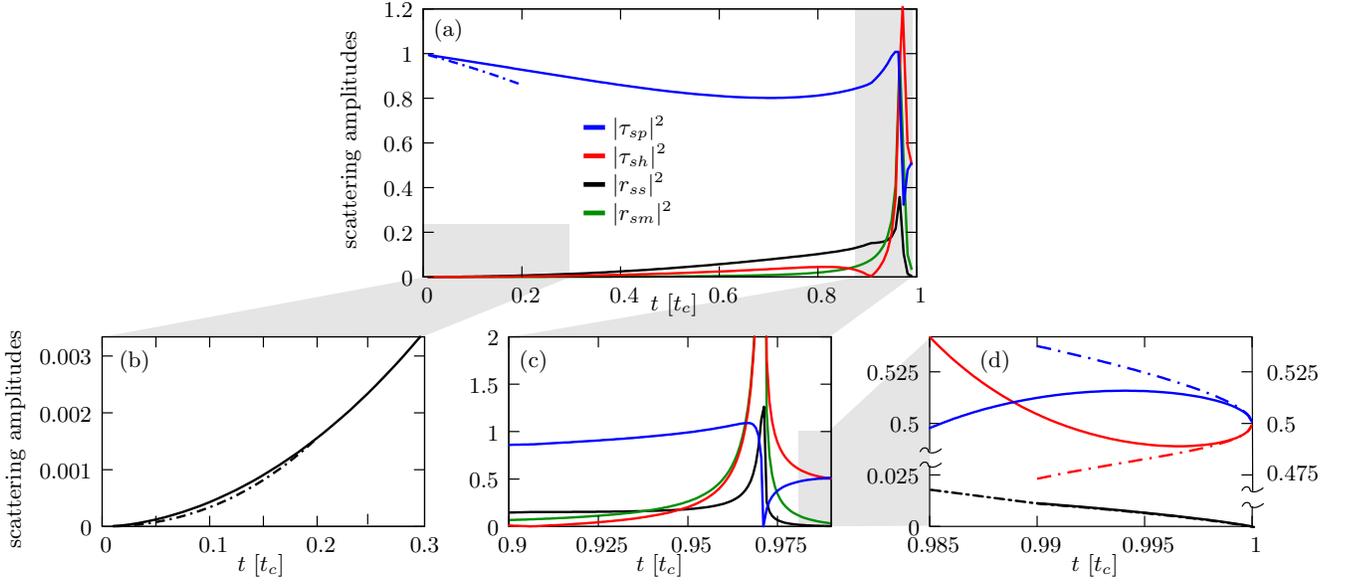}
\end{center}
\caption{
Scattering coefficients for an incoming sound mode at an energy
$\epsilon=\hbar\omega_{s}(\pi)/4$. (a) For all values of $t/t_{c}$.  The full
lines show the exact values, while the dash-dotted line is the approximate
value from Eq.\ (\ref{eqn:results-smallt}). (b) Comparison between the
exact and approximate values for $r_{ss}$. (c) All coefficients around the
``critical'' value $t\approx 0.971$ where for $\epsilon=\hbar\omega_{s}(\pi)
/4$ the hole and massive modes become exponentially damped.  Note, that
values of the scattering coefficients larger than 1 are not in contradiction
with particle conservation as these excitations do not have a well defined
particle character. (d) For values $t\approx t_{c}$ the dash-dotted results
from Eq.\ (\ref{eqn:results-largt}) again describe the exact solutions well.
}
\label{fig:scat_coff}
\end{figure*}

Next, we discuss a specific situation, the scattering of a phonon on
a Mott-insulator boundary, and determine the scattering coefficients
by matching the wave functions [Eq.\ (\ref{eqn:swf})] at the boundary
between the superfluid and the insulating region. The scattering state
(\ref{eqn:swf}) describes a phonon excitation incident from a particle-type
superfluid S$_p$ onto a Mott insulator [cf.\ Fig.~\ref{fig:overview}]. These
excitations are relevant in an experiment where a density perturbation is
applied to the middle of the trap;\cite{Kollath05} while one has to describe
the density perturbation by a wave packet of phonons, here, we only discuss
the scattering properties of plane-wave excitations as the generic building
block.  Imposing the continuity conditions across the S$_p$-MI interface,
we obtain the scattering coefficients $r_{ss}$, $r_{ms}$, $\tau_{ps}$,
and $\tau_{hs}$; only the first two spinor elements are relevant as the
second two satisfy the matching conditions automatically due to ${\mathcal
T}$ invariance. In the following, we discuss these coefficients in the
limiting cases $t\to 0,~t_c$ and illustrate their behavior at intermediate
values of $t$ in Fig.\ \ref{fig:scat_coff}.  Furthermore, here, we restrict
the discussion to those modes propagating in the superfluid as well as in
the Mott-insulator region; in the next section, where the heat transport
through a finite Mott region is discussed, the contribution of evanescent
modes has to be considered as well.

For $t\to 0$, a sound mode incident from S$_p$ can be reflected as a sound
mode or transmitted into the Mott insulator as a particle excitation;
hence. the coefficients connecting to propagating scattered modes are
$r_{ss}$ and $\tau_{ps}$; given the energy $\epsilon$ of the incident
phonon mode, they can be conveniently expressed through the wave vectors
$k_s^\epsilon$ and $k_p^\epsilon$ of the sound- and particle excitations
involved,
\begin{align}
\label{eqn:results-smallt}
  r_{ss}(\epsilon)
  &\approx i \frac {k_p^\epsilon-k_s^\epsilon} {k_s^\epsilon+k_p^\epsilon}
  + {\mathcal O}[( t/t_c)^2],
\\
\nonumber
  \tau_{ps}(\epsilon)
  &\approx -\frac{2k_s^\epsilon}{k_s^\epsilon+k_p^\epsilon}
  \biggl(1-\frac{t}{4t_c}\frac{\cos(k_s^\epsilon)}{1-\cos(k_s^\epsilon)}\biggr) 
  + {\mathcal O}[(t/t_c)^2]
\end{align}
(the coefficients $r_{ms}$ and $\tau_{hs}$ describe scattering into
evanescent modes).  To leading order in $t/t_c$, we recognize the standard
expressions for the one-dimensional barrier problem where the reflection and
transmission amplitudes can be expressed in terms of momentum ratios. This
result comes about since, for $t=0$, the two phases, S$_p$ and the Mott
insulator, are essentially identical.  However, for the current setup,
the height of the ``barrier'' is not a free parameter due to the generic
nature of the interface; the result then is determined by the dispersions
at the boundary. For a small but finite hopping $t/t_c$, the Mott insulator
remains essentially unchanged, while the bosons in the superfluid exploit
the kinetic energy. This leads to a reduced transmission due to the wave
function mismatch at the boundary, as reflected by the term $\!\propto
t/t_c$ in $\tau_{ps}$.

Next, we discuss the situation near the tip of the Mott lobe; in order
to formulate the results for $t\to t_c$, we first define the reduced
distance $\delta_t=1-t/t_c>0$ from the tip of the Mott lobe and the
coherence functions
\begin{equation}
	f_\pm(k)=[A(k)\pm i B(k)]\big|_{t=t_c}.
\end{equation}
While the phases $\arg[f_\pm(k)]$ run from $\arg[f_\pm(0)]=\mp \pi/4$
to $\arg[f_\pm(\pi)]=\pm {\rm acos} (1/ \sqrt{3}+ 1/\sqrt{6})$, both
moduli diverge at $k\to 0$, reflecting the softening of all modes at
the tip of the Mott lobe; however, the ratio $|f_+(k)/f_-(k)|^2\equiv1$
remains finite. Away from $k\rightarrow0$ the moduli of $f_{\pm}(k)$
are well behaved. To leading order in $\delta_{t}$ we obtain
\begin{align}
\label{eqn:results-largt}
r_{ss}(\epsilon) &\approx 
i \frac{k_p^\epsilon-k_s^\epsilon} {k_p^\epsilon+k_s^\epsilon} 
+ {\mathcal O}(\delta_t), 
\qquad  
r_{ms}(\epsilon) \approx {\mathcal O}(\delta_t), 
\\
\nonumber
\tau_{ps}(\epsilon) &\approx 
- \frac{\sqrt{2}k_s^\epsilon}{k_p^\epsilon+k_s^\epsilon}
   \biggl( \frac{f_-(k_s^\epsilon)}{f_+(k_p^\epsilon)}
   + \frac{f_-(k_s^\epsilon)}{f_-(k_p^\epsilon)}\sqrt{\delta_t} \biggr)
+ {\mathcal O}(\delta_t),
\\
\nonumber
\tau_{hs}(\epsilon) &\approx 
- \frac{\sqrt{2}k_s^\epsilon}{k_h^\epsilon+k_s^\epsilon}
   \biggl( \frac{f_-(k_s^\epsilon)}{f_+(k_h^\epsilon)}
   - \frac{f_-(k_s^\epsilon)}{f_-(k_h^\epsilon)}\sqrt{\delta_t} \biggr)
+ {\mathcal O}(\delta_t).
\end{align}
For $\delta_{t}\to 0$, we have $k_s^\epsilon \approx k_h^\epsilon \approx
k_p^\epsilon$, the coherence factors are approximately unity, and we again
recognize the typical result for a simple barrier. This time, however,
the particle and the hole branches are degenerate and are transmitted
equally to leading order, explaining the factor $\sqrt{2}$. Furthermore,
the massive mode is orthogonal to the sound mode and does not contribute
in leading order.  Going to nonzero $\delta_t$, the superfluid develops
``particle'' nature, leading to an increase (reduction) in $\tau_{ps}$
($\tau_{hs}$).

The behavior of the scattering coefficients for intermediate values of
$t/t_{c}$ is shown in Fig.\ \ref{fig:scat_coff}. The coefficients are
given for an energy $\epsilon=\hbar\omega_{s}(\pi)/4$ allowing for a
comparison at different values of $t/t_{c}$. Note that for this energy,
the hole and massive modes become non-propagating for $t/t_c\lesssim
0.971$.  The corresponding density of states effect is responsible for
the irregular behavior around $t/t_{c} \approx 0.971$. At small $t/t_{c}$,
the approximations [Eq.\ (\ref{eqn:results-smallt})] are in good agreement
with the exact values; the same is true near the tip of the Mott lobe,
where the coefficients [Eq.\ (\ref{eqn:results-largt})] agree well with
the exact numerical values.  Here, however, the close vicinity of the
point where the massive mode and the hole turn evanescent spoils the
applicability of the analytical results much faster.

Given the results above, we note the following generic trends: Scattering
coefficients relating modes of equal type, e.g., particle-type sound in
the S$_p$ and particle excitations in the Mott insulator are large, while
scattering between unequal modes, e.g., particle-type sound in the S$_p$
and hole excitations in the Mott insulator is suppressed. Furthermore,
the energy-dependent point on the $t/t_{c}$ axis where the massive and
hole modes turn evanescent controls the breakdown of the approximate
formulas (\ref{eqn:results-smallt}) and (\ref{eqn:results-largt}). The
same statements hold for the respective cases of an incoming massive mode
and the inverted situations at the lower boundary of the Mott lobe, i.e.,
with initial states in S$_h$.

\section{Heat conductivity}\label{sec:heatcond}

As an application, we calculate the transport of heat through a
Mott-insulating layer within a wedding cake structure. The derived heat
conductivity $\kappa$ provides insight into the thermalization process when
the lattice potential is ramped up,\cite{Pollet08} and quantifies to what
extent two superfluid shells are in thermal contact through the Mott layer.
In order to calculate $\kappa$, we consider the heat current\cite{Mahan-heat}
\begin{equation}
\label{eqn:heatcurrent}
{\bf Q}_{E}=\sum_{\alpha,\beta} \int d\epsilon\, 
N_{\alpha}(\epsilon)
{\bf v}_{\beta}(\epsilon) \epsilon 
g(\epsilon,T) |\tau_{\alpha\beta}|^{2},
\end{equation}
where the sum is running over input ($\alpha$) and output ($\beta$) channels.
In Eq.\ (\ref{eqn:heatcurrent}), $N_{\alpha}(\epsilon)$ is the density of
states in the input channels $\alpha$, ${\bf v}_{\beta}(\epsilon)$ is the
velocity associated with the excitations at energy $\epsilon$ in the output
channels $\beta$, and we integrate over all energies $\epsilon$. The
Bose-Einstein distribution $g(\epsilon,T) = [\exp(\epsilon/k_{\ssm
B}T)-1]^{-1}$ controls the occupation of the bosonic modes and
$\tau_{\alpha\beta}$ are the transmission amplitudes connecting input and
output channels. The linear-response expression Eq.\ (\ref{eqn:heatcurrent})
describes the situation close to equilibrium. In a real experiment, the
initial state after ramping of the optical potential may be far away from
the bosonic equilibrium distribution $g(\epsilon,T)$. Nevertheless, the
calculation of the linear-response result [Eq. (\ref{eqn:heatcurrent})]
provides some generic insights into the behavior of the system which
applies to such a situation as well.

It is instructive to calculate the maximal heat conductivity of a homogeneous
Mott region. For temperatures higher than all energy scales in the problem
(the repulsion $U$), but lower than the band gap to the next Bloch band,
the heat conductivity is finite and given by
\begin{align}
\nonumber
\kappa_{\infty}^{\ssm hom}&=
\sum_{\alpha=p,h} \int_{0}^{\pi/a}\!\! \frac{dk}{\pi/a}\,
v_{\alpha}(k) \hbar\omega_{\alpha}(k)
\partial_{T} g(\epsilon,T\gg U)\\
&=
\frac{Ua}{\hbar}k_{\ssm B}
\biggl(\sqrt{1+t/t_{c}}-\sqrt{1-t/t_{c}}\biggr),
\label{eqn:kappamax}
\end{align}
where we explicitly accounted for the lattice constant $a$.  From Eq.\
(\ref{eqn:kappamax}) we learn that the coefficient $\kappa$ describes the
transport of entropy ``$k_{\ssm B}$'' with velocity $Ua/\hbar$. Below,
we take
\begin{equation}
\kappa_{\ssm ref}
=\kappa_{\infty}^{\ssm hom}(t\rightarrow t_{c})
=\sqrt{2}\frac{Ua}{\hbar}k_{\ssm B}
\end{equation}
as our reference value for the heat conductivity.

\begin{figure}[!t]
\includegraphics{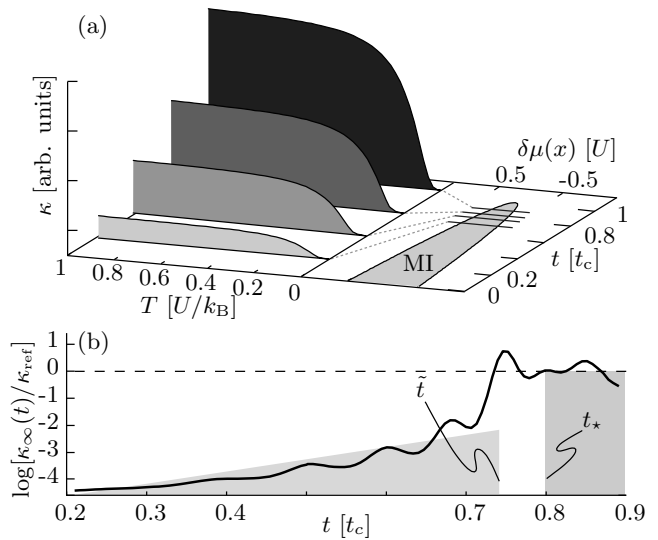}
\caption{
(a) Heat conductivity $\kappa(T)$ for different values of $t$
($t=0.9,0.82,0.78,0.74t_{c}$, from the back to the front) for a Mott
layer of thickness $L_{0}=10a$ at $t=0$ ($a$ denotes the lattice constant).
The structure of $\kappa(T)$ for small $T$ is dictated by the spectral gap in
the intermediate Mott region. For smaller values of $t$, where only damped
modes contribute to transport, the heat conductivity is exponentially
suppressed and not shown here. (b) The dependence of the maximal heat
conductivity $\kappa_{\infty}=\kappa(T\gg U/k_{\ssm B})$ on the hopping
amplitude $t$ displaying the qualitative change for $t<\tilde t$ in the plot,
where the strong suppression sets in. The change at $t_{\star}$ is masked
by a scattering resonance in the intermediate $\tilde t <t<t_{\star}$ regime.
\label{fig:kappa}
}
\end{figure}

We calculate the heat current [Eq.~(\ref{eqn:heatcurrent})] from one
superfluid shell, point $A$ in Fig.~\ref{fig:overview}(a), to point $B$
in the next shell and take the derivative with respect to the temperature
$T$ to obtain $\kappa(T)$ (cf.\ Fig~\ref{fig:kappa}). The input channels
$\alpha$ are given by the phonons and massive particles impinging onto the
Mott insulating phase at point $A$, i.e., from the particle-like superfluid
S$_p$. The output channels on the other side of the insulating region are
the corresponding modes in S$_h$.

To obtain the transmission amplitudes $\tau_{\alpha\beta}$, we apply
a transfer matrix formalism\cite{Azbel83} to divide the problem into
three parts, the scattering across the interface between the superfluid
S$_p$ and the Mott region, point $A$ in Fig.~\ref{fig:overview}(a),
the propagation within the Mott layer, and the scattering at the second
interface connecting the Mott insulator and the superfluid S$_h$ [cf.\
point $B$ in Fig.~\ref{fig:overview}(a)]. The scattering at the two
interfaces is handled as described in Sec.~\ref{sec:scatcoff} above; in
addition, scattering amplitudes into and from evanescent modes now have
to be accounted for.
\begin{figure*}[t]
\begin{center}
\includegraphics{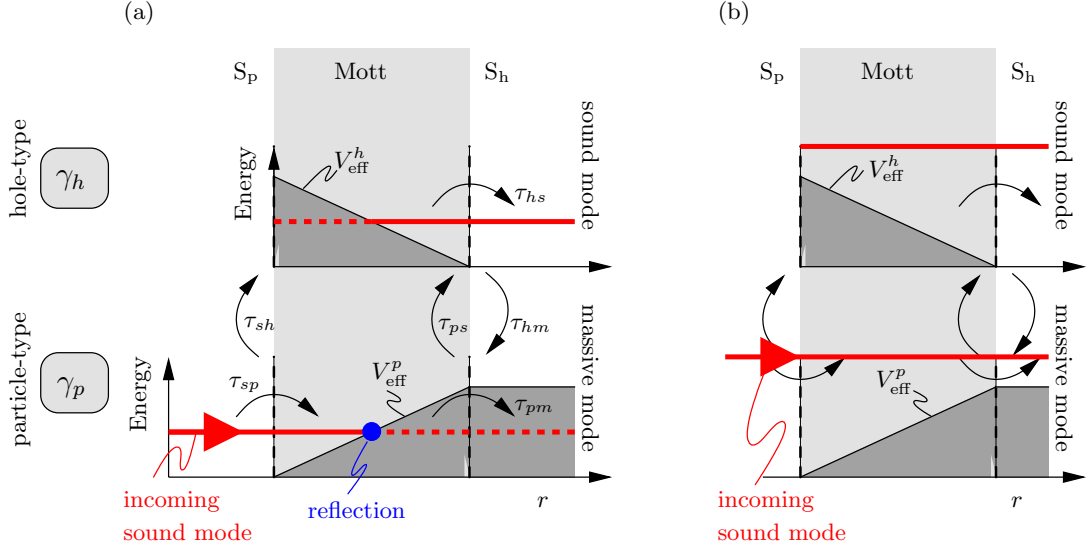}
\end{center}
\caption{
Sketch of a scattering event of an incoming phonon from S$_p$. (a)
The incident sound mode is transformed into a particle and a hole with
amplitudes $\tau_{sp}$ and $\tau_{sh}$, respectively.  Inside the Mott
layer, the particle propagates until it hits the potential, where it is
either reflected or tunnels under the barrier. The hole emerges from under
the barrier and propagates to the other end of the layer. At the boundary
to S$_h$, both modes are converted into the modes of the superfluid with
the corresponding amplitudes.  (b) The same situation with a higher energy
of the incoming sound mode that allows for undamped propagation through
the Mott region.
}
\label{fig:turn_point}
\end{figure*}

To describe the transfer from the superfluid phase into the Mott insulator
and vice versa, we match the wave functions [Eq.\ (\ref{eqn:wavefunctions})]
and their derivatives at the boundary (placed at $y=0$): we define the
wave functions
\begin{align}
\nonumber
\bfpsi_{\ssm MI}(y) \!&=\!
P_{r}  \bfpsi_{p}^{k}(y) \!+\!
P_{l} \bfpsi_{p}^{\!-k}(y) \!+\!
H_{r} \bfpsi_{h}^{k}(y) \!+\!
H_{l} \bfpsi_{h}^{\!-k}(y),
\\
\nonumber
\bfpsi_{\ssm S}(y) \!&=\!
S_{r} \bfpsi_{s}^{k}(y) \!+\!
S_{l}  \bfpsi_{s}^{\!-k}(y) \!+\!
M_{r} \bfpsi_{m}^{k}(y) \!+\!
M_{l} \bfpsi_{m}^{\!-k}(y),
\end{align}
and impose the matching conditions $\bfpsi_{\ssm MI}(0) = \bfpsi_{\ssm S}(0)$
and $ \partial_{y}\bfpsi_{\ssm MI}(0) = \partial_{y}\bfpsi_{\ssm S}(0)$ for
the two first components; the conditions for the other two components then
are fulfilled automatically due to ${\mathcal T}$-symmetry. The subscripts
$l$ and $r$ denote right- and left-moving excitations, respectively. This
procedure provides us with four relations connecting the four amplitudes
on one side of the interface with the four on the other.  Solving for
the coefficients $P_{l(r)}$ and $H_{l(r)}$, we obtain the transfer matrix
$\mathsf M_{\ssm S-MI}$ defined as
\begin{equation}
\label{eqn:newvectors}
\begin{pmatrix}
P_{r} \\
P_{l} \\
H_{r} \\
H_{l}
\end{pmatrix}
=
\mathsf M_{\ssm S-MI}
\begin{pmatrix}
S_{r} \\
S_{l} \\
M_{r} \\
M_{l}
\end{pmatrix}.
\end{equation}
This transfer matrix depends on the type of interface, S$_p$ or S$_h$,
connecting to the Mott layer; correspondingly, we denote the two
different matrices by $\mathsf M_{\ssm S-MI}^{\ssm p(h)}$. Note that the
four-dimensional character of Eq.\ (\ref{eqn:newvectors}) is due to the
restriction to the ``$+$''-sector of $\mathcal T$; in general, one expects
the transfer matrix to act on a vector space of twice the dimension of the
spinor. However, the matrix elements connecting the two sectors ``$+$''
and ``$-$'' vanish for a $\mathcal T$-symmetric system.

The new task to analyze is the propagation of particle- and hole excitations
through the Mott region $0<y<L$ as described by the transfer matrix
\begin{equation}
\begin{pmatrix}
P_{r}(y=L) \\
P_{l}(y=L)  \\
H_{r}(y=L)  \\
H_{l}(y=L)
\end{pmatrix}
=
\mathsf M_{\ssm MI}
\begin{pmatrix}
P_{r}(y=0)  \\
P_{l}(y=0)  \\
H_{r}(y=0)  \\
H_{l}(y=0)
\end{pmatrix}.
\end{equation}
This task requires the evaluation of the WKB-phases (between arbitrary
points $a$ and $b$)
\begin{equation}
\label{eqn:WKBphases}
\phi^{p(h)}_{a,b}=\int_{a}^{b} dx\, k_{p(h)}^{\epsilon}(x)
 \end{equation}
of Eq.\ (\ref{eqn:qcwf}) in the presence of an inhomogeneous potential
$V(y)$.  Attention has to be paied to properly treat the classical turning
points where the quasi-classical approximation [Eq.\ (\ref{eqn:qcwf})]
breaks down; Fig.\ \ref{fig:turn_point} illustrates typical situations
in the present context, where particles and holes incident from the left
are stopped by the potential $V_{\ssm eff}^{p(h)}$ and turn evanescent or
tunnel as evanescent modes into the MI and turn into propagating modes at
the right of the Mott region. Such turning points are dealt with in the
standard way\cite{Migdal77} and lead to additional scattering phases $\pm
\pi/4$ in the propagator.

\begin{figure*}[t]
\begin{center}
\includegraphics{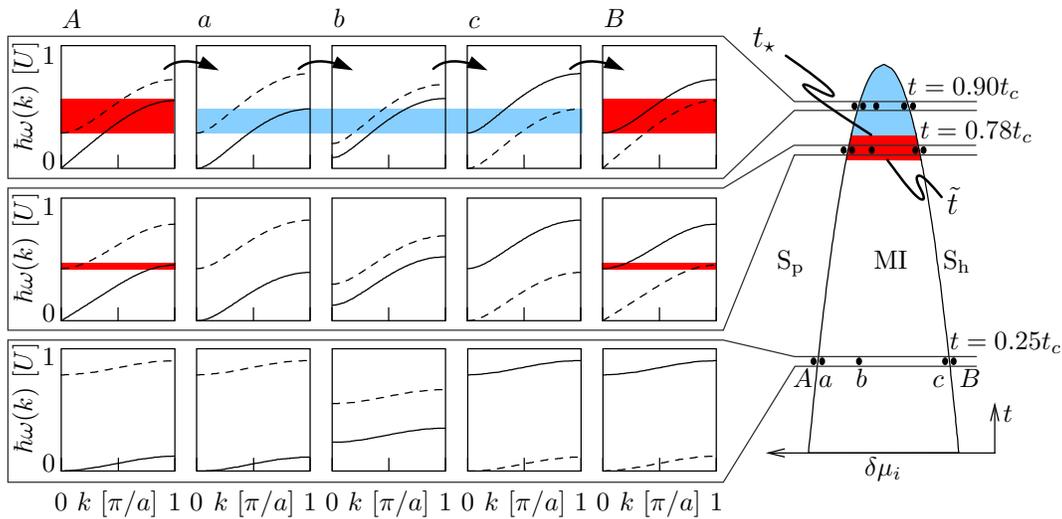}
\end{center}
\caption{
(color online) Left: sketch of the local excitation spectra at different
points $A,a,b,c,B$  (columns)  in the inhomogeneous system for three
different values of the hopping amplitude at $t/t_{c}=0.9,0.78,0.25$ (rows);
Hole-like spectra are dashed; particle-like spectra are solid lines. Right:
local phase diagram with Mott insulating and superfluid regions depending
on the chemical potential $\delta\mu_{i}$ and the hopping $t/t_{c}$. The
blue (light gray) shaded area of the Mott lobe corresponds to values of
$t/t_{c}$ for which the hole at point $a$ has spectral overlap with the
hole-like spectra at $c$, {\em allowing for undamped propagation} in the
energy window given in blue (light gray) on the left. The red (dark gray)
shaded area of the Mott lobe marks values of $t$ for which the modes in
the superfluid have mutual spectral overlap for an non-vanishing range of
energy as shown on the left. For values $t<\tilde t$, the {\em crossed
terms do not contribute} (phonon $\leftrightarrow$ massive mode) to the
heat conductivity (see text).
}
\label{fig:scenarios}
\end{figure*}

Depending on the appearance of turning points in the particle and/or hole
channel, the transfer matrix $M_{{\ssm MI}}$ assumes different forms. The
simplest case is realized near the tip of the Mott lobe, where particle
and hole modes can propagate unhindered through the Mott region [cf.\
Fig.\ \ref{fig:scenarios}].  In this case, the matrix is diagonal and
multiplies each component of the four-spinor with appropriate phase factors,
\begin{equation}
\nonumber
\mathsf{M}_{{\ssm MI}} = 
\begin{pmatrix} e^{i\varphi_{0,L}^{p}} & 0 & 0 & 0 \\ 0   &
e^{-i\varphi_{0,L}^{p}} & 0 & 0 \\
 0 & 0 &
e^{i\varphi_{0,L}^{ h}} & 0
 \\ 0 & 0 & 0 & e^{-i\varphi_{0,L}^{h}}
\end{pmatrix}=
\begin{pmatrix}
\mathsf{m}_{\ssm MI}^{p} &  0 \\
0 & \mathsf{m}_{\ssm MI}^{h}
\end{pmatrix}.
\end{equation}
The appearance of a turning point in the particle (hole) channel renders the
$2\times 2$ block matrix $\mathsf{m}_{{\ssm MI}}^p$ ($\mathsf{m}_{{\ssm
MI}}^h$) describing particle (hole) propagation non-diagonal; note
that particles (holes) then are reflected but never converted into
one another. Assuming a classical turning point at $r_{\ssm cl}$ (cf.\
trajectory $\gamma_{p}$ in Fig.\ \ref{fig:turn_point}), the transfer matrix
for a particle excitation assumes the form
\begin{equation*}
\mathsf m_{{\ssm MI}}^p =
\begin{pmatrix}
\frac{1}{2}e^{i\varphi_{0,L}^{p}-i\pi/4} &
\frac{1}{2}e^{-i\varphi_{0,r_{\ssm cl}}^{p}
              +i\varphi_{r_{\ssm cl},L}^{p}+i\pi/4} 
\\
e^{i\varphi_{0,r_{\ssm cl}}^{p}
  -i\varphi_{r_{\ssm cl},L}^{p}+i\pi/4} &
e^{-i\varphi_{0,L}^{p}-i\pi/4} 
\end{pmatrix}.
\end{equation*}
The equivalent expression for the hole excitation (trajectory $\gamma_{h}$
in Fig.\ \ref{fig:turn_point}), reads as
\begin{equation*}
\mathsf m_{{\ssm MI}}^h =
\begin{pmatrix}
e^{i\varphi_{0,L}^{h}+i\pi/4} &
\frac{1}{2} e^{-i\varphi_{0,r_{\ssm cl}}^{h}
               +i\varphi_{r_{\ssm cl},L}^{h}-i\pi/4} \\
e^{i\varphi_{0,r_{\ssm cl}}^{h}
  -i\varphi_{r_{\ssm cl},L}^{h}-i\pi/4} &
\frac{1}{2} e^{-i\varphi_{0,L}^{h}+i\pi/4}
\end{pmatrix}.
\end{equation*}
Note, that due to particle-hole symmetry, the turning points $r_{\ssm cl}$
are the same for particles and holes.  The ``phases'' $\phi_{a,b}^{p(h)}$
have to be calculated numerically. The total transfer matrix, connecting
the two superfluids on either side of the Mott insulator, is given by
\begin{equation}
\mathsf M_{\ssm tot} =
(\mathsf M_{\ssm S-MI}^{\ssm h})^{-1}
\mathsf M_{\ssm MI}
\mathsf M_{\ssm S-MI}^{\ssm p}.
\end{equation}
We obtain the scattering matrix $\mathsf S$ by solving the linear equation
\begin{equation}
\begin{pmatrix}
S_{\ssm out}^{\ssm h} \\
S_{\ssm in}^{\ssm h} \\
M_{\ssm out}^{\ssm h} \\
M_{\ssm in}^{\ssm h} \\
\end{pmatrix} =
\mathsf M_{\ssm tot}
\begin{pmatrix}
S_{\ssm in}^{\ssm p} \\
S_{\ssm out}^{\ssm p} \\
M_{\ssm in}^{\ssm p} \\
M_{\ssm out}^{\ssm p} \\
\end{pmatrix}
\end{equation}
for the amplitudes of the outgoing excitations $S_{\ssm out}^{\ssm p}$,
$M_{\ssm out}^{\ssm p}$, $S_{\ssm out}^{\ssm h}$, and $M_{\ssm out}^{\ssm
h}$. For the heat conductivity, we are only interested in the transmission
amplitudes connecting the channels of incoming phonons and massive modes
from S$_p$ to the outgoing excitations in S$_h$. Within this subspace the
scattering matrix reads as
\begin{equation}
\label{eqn:transmissionamp}
\begin{pmatrix}
S_{\ssm out}^{\ssm h} \\
M_{\ssm out}^{\ssm h}
\end{pmatrix}
=
\mathsf S
\begin{pmatrix}
S_{\ssm in}^{\ssm p} \\
M_{\ssm in}^{\ssm p}
\end{pmatrix}
=
\begin{pmatrix}
\tau_{ss} & \tau_{sm} \\
\tau_{ms} & \tau_{mm}
\end{pmatrix}
\begin{pmatrix}
S_{\ssm in}^{\ssm p} \\
M_{\ssm in}^{\ssm p}
\end{pmatrix},
\end{equation}
and provides us with the explicit form of the scattering amplitudes
$\tau_{\alpha\beta}$ appearing in the expression for the heat current [Eq.\
(\ref{eqn:heatcurrent})].

The density of states $N_\alpha(\epsilon)$ and velocity $v_\beta(\epsilon)$
derive from the dispersion relations $\omega_{s(m)}$ in Eq.\
(\ref{eqn:deltamass}). In one dimension, these quantities are related
via $v_{s(m)}=1/N_{s(m)}(\epsilon)$ and hence only in processes where
the phonon and massive mode channels are mixed with $\alpha\neq\beta$;
there appears a ratio $N_{s(m)}(\epsilon)/ N_{m(s)}(\epsilon)$, otherwise
all density of states effects cancel out.

Next, we identify those regions in the Mott lobe where we expect a large
value of $\kappa$.  Figure \ref{fig:scenarios} shows the evolution of the
spectra upon crossing the Mott-insulating region, starting with sound-
and massive modes in the particle-type superfluid S$_p$ at point $A$ right
at the interface, the swap of particle- and hole-type branches within the
MI regime with a potential rising approximately linearly with distance;
see diagrams ``a'', ``b'', and ``c'' in Fig.\ \ref{fig:scenarios}, and the
interchanged massive- and sound modes in the hole-type superfluid S$_p$
at point $B$, again right at the interface. Following the evolution of
the spectra from the tip of the Mott lobe at $t=t_c$ down to $t=0$,
various regimes can be identified where transport is favored, either
via full propagation through the Mott region or via conservation of
the particle/hole nature of the excitation along the trajectory. At
large values of $t$ close to the tip of the lobe an appreciable part
of the particle- and hole branches overlaps, allowing these excitations
to propagate through the Mott insulating without damping. This overlap
terminates when the bottom of the hole band lines up with the top of the
particle band [$\hbar\omega_{p}(\pi/a)=\hbar\omega_{h}(0)$; cf.\ Eq.\
(\ref{eqn:deltapart})  and diagram ``a'' in Fig.\ \ref{fig:scenarios}],
defining the special value $t_\ast=(4/5)t_{c}$. Another relevant point
is $\tilde t$, where the massive and sound modes stop overlapping in the
superfluid; see diagram ``$A$'' in Fig.\ \ref{fig:scenarios}.  Comparing
the bottom of the massive mode at $k=0$ with the top of the sound mode
at $k=\pi$, we find that this overlap persists as long as $t > \tilde t$
with $\tilde t = (3-\sqrt{5})t_{c}$; and we have $\Delta_{m} < W_{\varphi}$,
where $W_{s} = \hbar \omega_{s}(\pi/a)$ denotes the width of the sound mode
in S$_p$.  In this situation, the particle-type modes in the left (sound)
and right (massive) superfluids overlap (and vice versa for the hole modes)
and large transmissions at the boundaries enhance the contribution of these
modes to $\kappa$. Finally, for $t < \tilde t$, propagation through the
Mott region is always damped and excitations have to be converted between
particle- and hole-types, hence only direct terms $\propto |\tau_{ss}|^2$
(particle-type sound is converted to hole-type sound) and $\propto
|\tau_{mm}|^{2}$ (hole-type Higgs is converted to particle-type Higgs)
contribute. Accordingly, the heat conductivity $\kappa$ naturally splits
into direct and cross terms, with the latter ones contributing with large
weight but only for $t > \tilde t$ where $W_{s} > \Delta_{m}$,
\begin{equation}
\label{eqn:kappa}
\kappa(T)=\kappa_{\ssm dir}(T) + \Theta(W_{s}-\Delta_{m})
\kappa_{\ssm cross}(T),
\end{equation}
with
\begin{align}
\label{eqn:kappadir}
 \kappa_{\ssm dir}& =
\int_{0}^{W_{\varphi}}\!\!\!\!\!\!\!d\epsilon\,\epsilon |\tau_{ss}|^{2}
\frac{\partial g}{\partial T} +
\int_{\Delta_{m}}^{W_{m}}\!\!\!\!\!\!\!d\epsilon\,\epsilon |\tau_{mm}|^{2}
\frac{\partial g}{\partial T},\\
\label{eqn:kappacross}
\kappa_{\ssm cross}&=
\int_{\Delta_{m}}^{W_{\varphi}}\!\!\!\!\!\!\!d\epsilon\,\epsilon
\biggl[
\frac{N_{s}(\epsilon)}{N_{m}(\epsilon)}|\tau_{ms}|^{2}
+ \frac{N_{m}(\epsilon)}{N_{s}(\epsilon)}|\tau_{sm}|^{2}
\biggr]
\frac{\partial g}{\partial T}.
\end{align}
For the direct terms, the initial and the final states are both sound
modes or both massive modes and the density of states cancels against the
velocity factor.

The expressions in Eqs.\ (\ref{eqn:kappadir}) and (\ref{eqn:kappacross})
have to be calculated numerically and provide the final result for the
heat conductivity $\kappa(T)$ shown in Fig.\ \ref{fig:kappa} as a function
of $T/U$ for different values of $t/t_c$.  The overall shape involves an
exponential suppression at small temperatures $T$, hinting to the presence of
an effective gap $\Delta_{\ssm eff}$, and a saturation value $\kappa_\infty$
at large temperatures when all modes within the finite bandwidth contribute
to the transport. Furthermore, the transport efficiency decreases with
decreasing $t$, as is to be expected on the basis of the above analysis;
cf.\ Fig.\ \ref{fig:scenarios}.

Given the complexity of the result expressed through Eqs.\
(\ref{eqn:kappadir}) and (\ref{eqn:kappacross}) and the simplicity of
the final behavior of $\kappa$, we have attempted to extract a simple
and useful expression interpolating between the exponential rise and the
saturation at low and high temperatures. A phenomenological Ansatz with
a density of states $N(\epsilon) =\delta(\epsilon-\Delta_{\ssm eff})$
and a corresponding velocity $v_{0}$ provides us with the simple formula
\begin{equation}
  \kappa(T) \approx k_{\ssm B}v_{0}(\Delta_{\ssm eff})
  \left(\frac{\Delta_{\ssm eff}}{k_{\rm B}T}\right)^{2}
  \frac{\exp(\Delta_{\ssm eff}/k_{\rm B}T)}
  {[\exp(\Delta_{\ssm eff}/k_{\rm B}T)-1]^{2}},
  \label{eqn:phenomenkappa}
\end{equation}
with $x^2 \exp(x)/[\exp(x)-1]^2 |_{x\to 0} \to 1$, we find the velocity
$v_{0}$ parameter related to the saturation value of $\kappa$ at large
temperatures, $\kappa_{\infty} = k_{\ssm B}v_{0}$; the scaled function
$\kappa_{\infty} (t)$ shown in Fig.\ \ref{fig:kappa}(b) reproduces the
expected qualitative behavior, with a shoulder at $t\approx \tilde t$
and a suppression for $t<t_{\star}$.  The transmittance of the boundaries
as well as scattering resonances within the Mott layer (between the two
boundaries to the superfluids or between one such boundary and a classical
turning point) are all absorbed in the coefficient $\kappa_{\infty}$.
Also, we note that the band edges in the excitation spectra  manifest
themselves in the densities of states and give rise to sharp features
in $\kappa_{\infty}$.  These effects are masked in an experiment, where
the finite size of the superfluid and Mott-insulating regions induces an
uncertainty in the momenta of all excitations and we account for this
smearing in our numerical evaluation of Eqs.\ (\ref{eqn:kappadir}) and
(\ref{eqn:kappacross}).\cite{note:cutoff}

Close to the tip, where the conduction is dominated by itinerant modes,
the effective gap parameter $\Delta_{\ssm eff}$ turns out to be independent
of the Mott layer thickness $L$ and is approximately given by
\begin{equation}
\label{eqn:deltaeff}
\Delta_{\ssm eff}\approx t_{c}-t+0.32.
\end{equation}
The linear behavior in $t$ of $\Delta_{\ssm eff}$ has to be compared to the
evolution of the size of the maximal gap $\Delta_{m}$ within the Mott layer
which has a square root dependence on $1-t/t_{c}$. The different scaling
suggests that the gap $\Delta_{\ssm eff}$ plays the role of an effective
parameter describing the complex transport involving all of the interfaces
and the inhomogeneous Mott layer and is {\em not} directly related to the
spectral gap in the insulating region. In the same way, the offset by $\sim
0.32$ effectively accounts for the conversion of quasi-particles at the
phase boundaries.  With the above choices for the two phenomenological
parameters $v_{0}$ and $\Delta_{\ssm eff}$, we find excellent agreement
between the numerical data and the results of our simple Ansatz. In the
regime $t<\tilde t$, the effective gap depends on the thickness $L_{0}$
of the Mott layer ($L_{0}$ is taken at $t=0$). In Fig.~\ref{fig:kappa},
we show the results for $L_{0}=10a$. The slope of the exponential decrease
of $\kappa_{\infty}$ for $t<\tilde t$ depends strongly on $L_{0}$. However,
its significant suppression at $L_{0}=10a$ indicates that already at this
width, the superfluid layers are essentially decoupled even at temperatures
$k_{\ssm B}T\approx U$.

\section{Conclusions}
\label{sec:conclusions}

Summarizing, we have developed a framework to address the behavior of
quasi-particle excitations in a strongly correlated bosonic heterostructure.
We have derived a set of first-quantized spinor wave functions valid in
the superfluid- and Mott-insulating phases, have derived the scattering
properties of a superfluid-Mott-insulator interface, and have calculated
the heat conductance across a Mott region as an application of our method.

For a phonon incident from a S$_p$ onto a Mott insulator, we find standard
expressions for the scattering amplitudes describing the scattering at
a potential barrier in the limits $t\to 0,~t_{c}$, with the involved
momenta determined by the non-trivial bulk dispersions. Going away from
the Mott-lobe base and tip, the amplitudes pick up non-trivial corrections
which are easily found numerically.

In calculating the heat conductivity across a Mott layer in a wedding cake
structure, we have combined the scattering at the two interfaces with the
quasi-classical propagation through the inhomogeneous Mott layer. We find
that for a Mott shell at moderately to large hopping, i.e., $t>0.8t_{c}$,
the adjacent superfluid shells are in good thermal contact. For small
hopping below $\tilde t$, however, the Mott shells represent practically
infinite barriers.  Implications of our findings on the lattice ramping
problem\cite{Pollet08} deserve further studies.

\begin{acknowledgments}
We thank F.\ Hassler and L.\ Pollet for extensive discussions and acknowledge
financial support from the Swiss National Foundation through the NCCR MaNEP.
S.D.H. acknowledges the hospitality of the Institute Henri Poincar\'e--Centre
Emile Borel.
\end{acknowledgments}


\begin{thebibliography}{22}
\bibitem[{not({\natexlab{a}})}]{note:josephson}
\bibinfo{note}{This would not be the case for a discussion of the Josephson
  effect.}
\bibitem[{not({\natexlab{b}})}]{note:cutoff}
\bibinfo{note}{In order to account for this smearing, we cut off the integrals
  at low momenta $k_{c} \propto 1/L_{0}$, and add a convolution to the
  integrals in Eq.~(\ref{eqn:kappadir}) and (\ref{eqn:kappacross}) of the form
  $\int d\epsilon \rightarrow \int d\epsilon d\epsilon' f(\epsilon-\epsilon')$
  where $f(x)$ is a Lorentzian which is centered at $x=0$ and has a width
  $\propto 1/L_{0}^{2}$.}
\end{thebibliography}
\end{document}